\documentstyle[aps,prb]{revtex}
\def\b{\bibitem}
\def\be{\begin{equation}}
\def\ee{\end{equation}}
\def\bea{\begin{eqnarray}}
\def\eea{\end{eqnarray}}
\def\bml{\begin{mathletters}}
\def\eml{\end{mathletters}}
\begin{document}
%\makeatletter
\ifpreprintsty \else
\twocolumn[\hsize\textwidth\columnwidth\hsize\csname@twocolumnfalse%
\endcsname\fi

\title{The quantum phase transition of itinerant helimagnets}
\author{Thomas Vojta and Rastko Sknepnek}
\address{Institut f{\"u}r Physik, Technische Universit\"at Chemnitz, D-09107 Chemnitz, Germany}
\date{\today}
\maketitle
\begin{abstract}
We investigate the quantum phase transition of itinerant electrons from a
paramagnet to a state which
displays long-period helical structures due to a Dzyaloshinskii instability
of the ferromagnetic state.
In particular, we study how the self-generated effective long-range
interaction recently
identified in itinerant quantum ferromagnets is cut-off by the helical
ordering. We find that for a sufficiently strong Dzyaloshinskii instability
the helimagnetic
quantum phase transition is of second order with mean-field
exponents. In contrast, for a weak Dzyaloshinskii instability
the transition is analogous to that in itinerant quantum
ferromagnets, i.e. it is of first order, as has been observed in MnSi.
\end{abstract}

\ifpreprintsty \else
] \fi      % of \twocolumnfalse
%\narrowtext

\section{Introduction}
Quantum phase transitions are phase transitions that occur at zero temperature
as a function of some non-thermal control parameter like pressure, magnetic
field, or chemical composition. While the usual finite-temperature phase
transitions
are driven by thermal fluctuations, zero-temperature quantum phase transitions
are driven by quantum fluctuations which are a consequence of Heisenberg's
uncertainty principle. Quantum phase transitions have
attracted considerable attention in recent years, in particular since they are
believed to be at the heart of some of the most exciting discoveries in modern
condensed matter physics, such as the localization problem, the quantum Hall
effects, various magnetic phenomena, and high-temperature
superconductivity.\cite{sgcs97,zhang97,kb97,sachdev00,vojta2000}

One of the most obvious examples of a quantum phase transition
is the transition from a paramagnetic to a ferromagnetic metal
that occurs as a function of the exchange coupling between the
electron spins. In a pioneering paper this transition was studied
by Hertz\cite{hertz76} who generalized Wilson's renormalization
group to quantum phase transitions. The finite temperature properties
were later discussed by Millis.\cite{millis93}
Building on these results the theory of the ferromagnetic quantum phase
transition has recently been worked out in much detail.
It was shown that in a zero-temperature correlated itinerant electron system
additional non-critical soft modes couple to the order
parameter. This effect produces a (self-generated) effective long-range
interaction between the spin fluctuations, even if the microscopic
exchange interaction is short-ranged.\cite{us_fm}
In a clean system the resulting ferromagnetic quantum phase
transition is generically of first order.\cite{us_first}

The experimentally best studied example of such a transition
is probably provided by the pressure-tuned transition in MnSi.\cite{PML95,PMJL97}
MnSi belongs to the class of nearly or weakly ferromagnetic metals.
These materials are characterized by strongly enhanced spin fluctuations.
Thus, their ground state is close to a ferromagnetic instability which
makes them good
candidates for actually reaching the ferromagnetic quantum phase transition
experimentally.
At ambient pressure MnSi is paramagnetic for temperatures larger
than $T_c=30$\,K. Below $T_c$ it orders magnetically. The phase transition
temperature can be reduced by applying pressure, and at about 14 kBar the
magnetic phase vanishes altogether. Thus, at 14 kBar MnSi undergoes a
magnetic quantum phase transition. The properties of this transition
are in semi-quantitative agreement with the theoretical
predictions\cite{us_first}, in particular,
the quantum phase transition is of first order while the thermal transition at
higher temperatures is of second order.\cite{FDFOT}

However, the magnetic order in MnSi is not exactly ferromagnetic but a
long-wavelength (190\,\AA)  helical spin spiral along the (111) direction of the
crystal. The ordering wavelength depends only weakly on
the temperature, but a homogeneous magnetic field of about 0.6\,T suppresses
the spiral and leads to ferromagnetic order. The helical structure
is a consequence of the so-called Dzyaloshinskii
mechanism,\cite{dzyaloshinskii64,bj80} an instability of the ferromagnetic
state with respect to small `relativistic' spin-lattice or spin-spin
interactions.
The helical ordering in MnSi immediately leads to the question, to what extent
the properties of the quantum phase transition in MnSi are generic for
itinerant quantum ferromagnets or whether the agreement between the
experiments and the ferromagnetic theory is accidental.

In this paper we therefore study how the long-period helimagnetism
caused by a Dzyaloshinskii instability influences the properties of the
quantum phase transition of an itinerant magnet.
We show that the self-generated  long-range interaction
between the spin fluctuations is cut-off by the helical
ordering. Depending on the relative strengths of this interaction and the
Dzyaloshinskii instability two different quantum phase transition scenarios
are possible. For a sufficiently strong Dzyaloshinskii instability the
quantum phase transition is of second order with mean-field
exponents while for a weak Dzyaloshinskii instability
the transition remains first order with the same properties as the
quantum ferromagnetic transition. This is the case realized in MnSi.

The paper is organized as follows. In Sec.\ \ref{sec:IQHM} we define the
starting action and identify the helical ordering at mean-field level.
In Sec.\ \ref{sec:QPT} we derive an order parameter field theory for
the paramagnet--helimagnet quantum phase transition, analyze it by
means of the renormalization group and discuss the resulting phase transition
scenarios. We conclude in Sec.\ \ref{sec:CONCL}.

\section{Itinerant quantum helimagnets}
\label{sec:IQHM}
\subsection{The model}
\label{subsec:MODEL}
Our starting point is the effective action for the spin degrees of
freedom in a three-dimensional itinerant quantum ferromagnet. This action
can be derived from a microscopic model of interacting electrons.\cite{us_fm}
In terms of the magnetization ${\bf M}$ the action reads
\begin{eqnarray}
\label{eq:fm_action}
S_{\rm FM}[{\bf M}]&=& \frac 1 2 \int dx dy\, {\bf M}(x)\,\Gamma_0(x-y)\,{\bf M}(y) \\
          &+& \sum_{n=3}^{\infty} a_n \int dx_1 \ldots dx_n\
              \chi^{(n)}(x_1,\ldots,x_n) \times  \nonumber\\
          & & \hspace*{80pt} \times\, {\bf M}(x_1) \ldots {\bf M}(x_n) ~.
               \nonumber
\end{eqnarray}
We have used a 4-vector notation with $x=({\bf r},\tau)$
comprising a real space vector ${\bf r}$ and imaginary time $\tau$.
Analogously, $\int dx = \int d{\bf r} \int_0^{1/T} d\tau$, where $T$
is the temperature.
The bare Gaussian vertex $\Gamma_0$ is proportional to $(1 - J \chi^{(2)})$ where
$J$ is the spin-triplet (exchange) interaction amplitude and $\chi^{2}$
is the spin susceptibility of a reference system which is a Fermi liquid
(precisely, it is the original electron system with the bare spin-triplet
interaction taken out).
In the appropriate limit of small momenta and frequencies the Fourier
transform of $\Gamma_0$ is given by
\begin{equation}
\Gamma_0({\bf k},\omega)=t_0 + B_1 k^2 + C_3 k^2 \log (1/k) +
C_\omega \frac {|\omega|} k~.
\label{eq:vertex}
\end{equation}
Here the third term represents the effective long-range interaction
induced by the coupling between the magnetization and non-critical soft
modes. Generically, it is repulsive, i.e $C_3<0$, but rather weak,
$|C_3|\ll B_1$ since it is caused by electronic correlations.
Therefore it becomes important only at small momenta. The last term
is the conventional Landau damping of the paramagnon.
The coefficients $\chi^{(n)}$ of the higher order terms in eq.\
(\ref{eq:fm_action})
are proportional to the higher spin density correlation functions of the
reference system. Because of the same mode coupling effects that lead to the non-analytic $C_3$ term
in the Gaussian action they are in general not finite in the limit of zero
frequencies and wave numbers. For ${\bf p} \to 0$ they behave like
$\chi^{(n)} \sim v^{(n)} |{\bf p}|^{4-n}$.

We now add a new term, the helical or Dzyaloshinskii
term,\cite{dzyaloshinskii64,bj80} to the effective action
(\ref{eq:fm_action}), which will cause an instability of the
ferromagnetic state,
\begin{equation}
S[{\bf M}]=S_{\rm FM}[{\bf M}] + D \int dx \, {\bf M}(x) \cdot {\rm curl}\,
{\bf M}(x)~.
\label{eq:full_action}
\end{equation}
Physically, this term may be caused by relativistic interactions between
spins of the form ${\bf S}_i \times {\bf S}_j$.
In general, it will be small compared to the other Gaussian terms,
with the possible exception of the long-range $C_3$ term.

\subsection{Mean-field theory: The spiral ordering}
\label{subsec:MF}
In this subsection we analyze the action (\ref{eq:full_action}) at
saddle point-level, i.e we minimize it with respect to ${\bf M}(x)$.
In the absence of the Dzyaloshinskii term the action is minimized by
a homogeneous (in space and imaginary time) ${\bf M}(x)$.
If the Dzyaloshinskii term is present, the action is minimized by
a state which is periodic in space but homogeneous in time:
\begin{equation}
{\bf M}({\bf r}, \tau)= {\bf A_k} e^{i {\bf k \cdot r}}
                      + {\bf A^*_k} e^{-i {\bf k \cdot r}}~.
\end{equation}
Here ${\bf A_k}={\bf a_k} + i {\bf b_k}$ is a complex vector.
Inserting this ansatz into the action (\ref{eq:full_action}) we obtain
\begin{eqnarray}
S^{\rm SP}({\bf k})&=& \left[t_0+ B_1 k^2 +C_3 k^2 \log (1/k)
 \right] |{\bf A_k}|^2 \nonumber \\
&&+ i D\, {\bf k} \cdot ({\bf A_k} \times {\bf A_k^*}) + O(|{\bf A_k}|^4)~.
\end{eqnarray}
The Gaussian part of $S^{\rm SP}({\bf k})$ is minimized for
$|{\bf a_k}| = |{\bf b_k}|$ and ${\bf a_k} \perp {\bf b_k}$.
The sign of $D$ determines the handedness of the resulting spin spiral.
For $D>0$ the minimum action is achieved for ${\bf k}$ antiparallel to
${\bf a_k} \times {\bf b_k}$, this is a right-handed spiral.  In contrast,
for $D<0$ the vector ${\bf k}$ must be parallel to
${\bf a_k} \times {\bf b_k}$, leading to a left-handed spiral.
Taking all these conditions into account the saddle-point action
reads
\begin{eqnarray}
S^{\rm SP}({\bf k})&=& \left[t_0+ B_1 k^2 +C_3 k^2 \log (1/k) - 2 |D| k
 \right] |{\bf A_k}|^2 \nonumber \\
&&+ O(|{\bf A_k}|^4)~.
\label{eq:SP}
\end{eqnarray}
The term in brackets is minimized by the ordering wave vector
${\bf K}$. Since in general $|D| \ll B_1$ the ordering wavevector
will be small. The direction of ${\bf K}$ cannot be determined
from our rotational invariant Gaussian vertex (\ref{eq:vertex}).
It will be fixed by additional (weak) anisotropic terms in the
model. We will come back to these terms in the next subsection.
In MnSi the spiral wavevector is known to be parallel to the (111)
or equivalent crystal directions.\cite{PML95,PMJL97,spiral} In the
following we will focus on this case, a
generalization to other directions is straightforward.

\section{The helimagnetic quantum phase transition}
\label{sec:QPT}
\subsection{Order parameter field theory}
\label{subsec:OPFT}

In the last subsection we found that the action is minimized by
long-period spin spirals which in MnSi are parallel to the (111)
direction. There are four equivalent ordering wave vectors ${\bf K}_j$,
{\it viz.} $K\,(1,1,1)/\sqrt{3}$, $K\,(-1,-1,1)/\sqrt{3}$,
$K\,(-1,1,-1)/\sqrt{3}$, and $K\,(1,-1,-1)/\sqrt{3}$.
For each wave vector ${\bf K}_j$ there are two equivalent directions
in the plane orthogonal to ${\bf K}_j$. Together this defines
8 equivalent spirals, i.e. the order parameter has eight components,
$\psi_j, \bar{\psi}_j$, $(j=1\ldots4)$.\cite{bj80}
We now consider slow fluctuations of the order parameter by writing the
magnetization as
\begin{eqnarray}
\label{eq:M-eta}
{\bf M}({\bf r},\tau)= \sum_{j=1}^4 &&~\left\{
        \psi_j({\bf r},\tau)
        \left[   {\bf A}_{{\bf K}_j} e^{i {\bf K}_j \cdot {\bf r}}
               + {\bf A}^*_{{\bf K}_j} e^{-i {\bf K}_j \cdot {\bf r}} \right]\right.\\
    &&\left.   + \bar{\psi}_j({\bf r},\tau)
        \left[   -i{\bf A}_{{\bf K}_j} e^{i {\bf K}_j \cdot {\bf r}}
               + i{\bf A}^*_{{\bf K}_j} e^{-i {\bf K}_j \cdot {\bf r}} \right]
               \right\},\nonumber
\end{eqnarray}
where ${\bf A}_{{\bf K}_j}={\bf a}_{{\bf K}_j} + i {\bf b}_{{\bf K}_j}$ with
$|{\bf a}_{{\bf K}_j}| = |{\bf b}_{{\bf K}_j}|=1$, ${\bf a}_{{\bf K}_j}
\perp {\bf b}_{{\bf K}_j}$, and ${\bf K}_j$
parallel or antiparallel to ${\bf a}_{{\bf K}_j} \times {\bf b}_{{\bf K}_j}$
depending on the sign of $D$.
The order parameter fields $\psi_j({\bf r},\tau)$ and
$\bar{\psi}_j({\bf r},\tau)$ are slowly varying in space and imaginary time,
in particular, they are only slowly varying over the wavelength of the spiral.
Inserting the order parameter representation (\ref{eq:M-eta}) into the action
(\ref{eq:full_action}) leads to the desired order parameter field theory
for the itinerant quantum helimagnet. The leading terms in an expansion
of the Landau-Ginzburg-Wilson free energy functional $\Phi$
in powers of momenta and frequencies of the order parameter field are
given by
\begin{eqnarray}
\Phi[\psi_j,\bar{\psi}_j]&=& \frac 1 2 \sum_{{\bf q},\omega,j}
    \left[|\psi_j({\bf q},\omega)|^2 + |\bar{\psi}_j({\bf q},\omega)|^2 \right ] \times
\nonumber \\
  &&\times \left[ t + B_1 q^2 + C_3 \,K^2 \log \left(\frac 1 K\right) +  \right. \nonumber
\\
 && + \left.  \tilde{B}_1 q^2 \log(1/K) +
     C_\omega \frac {|\omega|} K + O(q^3,\omega q^2) \right]\nonumber
\\
&+& O(\psi^4,\bar{\psi}^4)~.
\label{eq:LGW}
\end{eqnarray}
As a consequence of the spiral magnetic ordering the non-analyticities in the
Gaussian vertex (\ref{eq:vertex}) are cut-off at the ordering wave vector
${\bf K}$. For clarity we have written the resulting
$K$-dependent terms explicitly
in (\ref{eq:LGW}). In the following they will be absorbed into $t$ and
$B_1$, respectively. Analogously, in the frequency-dependent term $K$ will
be absorbed into $C_\omega$. The spiral ordering cuts-off not only the
non-analyticities in the Gaussian vertex but also the singularities
in the higher-order terms. Therefore the coefficients of all higher-order
terms in (\ref{eq:LGW}) are now finite in the limit ${\bf q},\omega  \to 0$.
Keeping only the most relevant terms (in the renormalization group sense)
and suppressing unessential constants,
the Landau-Ginzburg-Wilson functional $\Phi$ can finally be written as
\begin{eqnarray}
\Phi[\psi_j,\bar{\psi}_j]&=& \frac 1 2 \sum_{{\bf q},\omega,j} (t+q^2+|\omega|)
    \left[|\psi_j({\bf q},\omega)|^2 + |\bar{\psi}_j({\bf q},\omega)|^2 \right ]
    \nonumber
\\&+& u\int dx \left \{\sum_{j} [\psi_j^2(x) +\bar{\psi}_j^2(x)]\right\}^2
\nonumber
\\
&+& \lambda \int dx \sum_{j}\, [\psi_j^2(x) +\bar{\psi}_j^2(x)]^2~.
\label{eq:LGW-II}
\end{eqnarray}
Here the $u$ term is the conventional isotropic 4th order term, while
the $\lambda$ term represents a cubic anisotropy connected with the
discrete 4-fold degeneracy of the action with respect to the direction
of the spiral wave vector ${\bf K}$. One might worry whether additional
relevant contributions to (\ref{eq:LGW-II}) arise from the anisotropic terms
in the action necessary to fix
the directions of the spirals, as discussed after (\ref{eq:SP}). However,
once the rotational symmetry is broken by the discrete set of spiral directions
additional anisotropic terms in the action do not produce new contributions
to (\ref{eq:LGW-II}). An explicit calculation shows that they
only renormalize the coefficients $u$ and $\lambda$.

\subsection{Renormalization group analysis}
\label{subsec:RG}
We now analyze the Landau-Ginzburg-Wilson free energy functional
(\ref{eq:LGW-II}) by conventional
renormalization group methods for quantum phase
transitions.\cite{hertz76} We first carry out
a tree-level analysis. The Gaussian fixed point is defined by the requirement
that the coefficients of the $q^2$ and $|\omega|$ terms in the
Gaussian vertex do not change
under renormalization. Therefore, the dynamical exponent is $z=2$.
The other critical exponents which can be read off the Gaussian vertex
take their mean-field values: $\nu=1/2$, $\gamma=1$, and $\eta=0$.
Defining the scale dimension of a length to be $[L]=-1$, we find
the scale dimension of the fields at the Gaussian fixed point
to be $[\psi]=[\bar{\psi}]= (d+z-2)/2=3/2$
($d$ is the spatial dimensionality)

The properties of the Gaussian fixed point in our model
are identical to those of a conventional itinerant antiferromagnet.
This is not surprising since the structure of the Gaussian vertex
of the Landau-Ginzburg-Wilson functional (\ref{eq:LGW-II}) is identical
to that derived by Hertz\cite{hertz76} for itinerant antiferromagnets.
The only difference is the number of order parameter components.

In order to check the stability of the Gaussian fixed point we calculate the
scale dimensions of the coefficients of the quartic terms, $u$ and $\lambda$.
They turn out to be $[u]=[\lambda]=d+z-4[\psi]=4-d-z$. In our case, $d=3$, $z=2$
this means $[u]=[\lambda]=-1$. The quartic terms are irrelevant
at the Gaussian fixed point which is therefore stable. This is again
analogous to a conventional itinerant quantum antiferromagnet, the more
complicated order parameter component structure does not play any role
at the Gaussian fixed point.
Consequently, the helimagnetic quantum-critical point, if any, is characterized
by the usual mean-field exponents and a dynamic exponent of $z=2$.

\subsection{Phase transition scenarios}
\label{subsec:Scenarios}
In the last two subsections we have seen that in the long-distance and long-time
limit the order parameter field theory takes the form of an itinerant
quantum antiferromagnet, leading to a continuous transition with
mean-field exponents. However, in order to discuss all possible
scenarios for the helimagnetic quantum phase transition it is necessary to keep
track of the physics at finite length scales.
To do this it is useful to go back to the action (\ref{eq:full_action}) in
terms of the spin variables. In addition to the magnetic correlation length $\xi$ the
system is characterized by two non-trivial length scales, {\it viz.} the
length scale of the spiral, $\ell_{\rm Spiral}=|{\bf K}|^{-1}$, and the nucleation
length scale $\ell_{\rm Nucl}$ associated with the first-order transition in
itinerant quantum ferromagnets.\cite{us_first}
It is given by the length at which the $B_1$ and $C_3$ terms in the Gaussian
vertex (\ref{eq:vertex}) are equal and opposite, i.e.
$\log \, \ell_{\rm Nucl}=B_1/|C_3|$.

The qualitative properties of the helimagnetic quantum phase transition
are determined by the relation between these two length scales. Let us first
consider the case $\ell_{\rm Spiral} \ll \ell_{\rm Nucl}$, i.e.\ a strong
Dzyaloshinskii instability or weak electronic correlations.
When approaching the transition the growing
magnetic correlation length first reaches $\ell_{\rm Spiral}$. The system crosses over
from a ferromagnet to a helimagnet {\em before} the
self-generated effective long-range interaction becomes sufficiently strong
to induce a first-order transition. In this case the asymptotic action is given
by (\ref{eq:LGW-II}). The helimagnetic quantum phase transition is therefore
a continuous transition with mean-field critical exponents.

In the opposite case, $\ell_{\rm Spiral} \gg \ell_{\rm Nucl}$, i.e.\ a weak
Dzyaloshinskii instability or strong electronic correlations, the magnetic
correlation
length reaches $\ell_{\rm Nucl}$, before the Dzyaloshinskii term becomes
strong enough to induce spiral ordering. The system undergoes a first-order
phase transition which is completely analogous to the ferromagnetic transition.
Here, the correlation length remains finite (of the order of $\ell_{\rm
Nucl}$), and the long-wavelength free energy functional
(\ref{eq:LGW-II}) does {\em not} describe the transition.

In order to describe the crossover between the two scenarios we derive
a mean-field free energy for the helimagnetic quantum phase transition.
A mean-field theory is sufficient to obtain the leading behavior in both
scenarios: for the continuous transition this was shown in Subsec.\
\ref{subsec:RG} and for the first-order scenario in Ref. \onlinecite{us_first}.
In the ordered phase the non-analyticity in the long-range interaction in
the Gaussian vertex (\ref{eq:vertex}) is cut off\cite{us_first} by both a finite
order parameter $\psi$ and by the spiral wave vector ${\bf K}$.\cite{reason}
Taking both cutoffs into account we obtain
\begin{eqnarray}
F&=&t\,\psi^2 - C_3\,\psi^4\,\ln(\psi^2 + K^2) + \tilde{u}\,\psi^4 + O(\psi^6)~.
\label{eq:mf_free_energy}
\end{eqnarray}
For small $K$ and large $C_3$ this free energy displays a first order
transition,\cite{us_first} in the opposite case a continuous transition.
There is a quantum tricritical point at $|K|=\exp(-\tilde{u}/2|C_3|)$ which separates
the two regimes. The mean-field theory (\ref{eq:mf_free_energy}) also correctly
describes the quantum tricritical behavior at this point which is conventional.

\section{Conclusions}
\label{sec:CONCL}
In this final section of the paper we relate our findings to the
experiments on the quantum phase transition in the prototypical
itinerant helimagnet, MnSi.\cite{PML95,PMJL97}

In subsection \ref{subsec:Scenarios} we found that the properties of the helimagnetic
quantum phase transition crucially depend on the ratio of two length
scales, {\it viz.}
the wave length $\ell_{\rm Spiral}$ of the spiral and the nucleation
length $\ell_{\rm Nucl}$ associated with the first-order transition in
the corresponding itinerant quantum ferromagnet.
In MnSi the wave length
of the spiral is rather large, approximately 190\,\AA. In contrast, the
experimental data for the magnetic susceptibility suggest that the
nucleation length of the first order transition is small (of the order of
the microscopic scales). This can be seen from the fact that no susceptibility
increase is observed close to the quantum phase transition, instead the
susceptibility close to the transition is approximately a step function.
(If the first order transition would occur at some large length scale
the susceptibility should increase when approaching the transition
until the magnetic correlation length reaches this scale.)

Therefore, the nucleation length scale is much shorter than the
spiral wave length, and Subsection \ref{subsec:Scenarios}
predicts a first-order transition, in agreement with the experiments.
According to our theory the properties of the quantum phase transition in MnSi
are identical to that of the quantum {\em ferro}magnetic transition and MnSi
is indeed a prototypical example for this transition.

In summary, we have studied the quantum phase transition of itinerant
electrons from a paramagnet to a state which
displays long-period helical structures due to a Dzyaloshinskii instability
of the ferromagnetic state.
We found that depending on the relative strengths of the
helical (Dzyaloshinskii) term and the correlation-induced self-generated
long range interaction two different phase transition
scenarios are possible.
If the self-generated long-range interaction is stronger than the
helical term the transition is of first-order
with the same properties as the quantum ferromagnetic transition.
This is the situation encountered in MnSi.
In contrast, if the helical term is stronger the transition is a continuous
one with mean-field critical exponents and a dynamical exponent of $z=2$.
The two regimes are separated by a quantum tricritical point.

We gratefully acknowledge helpful discussions with S. Bekhechi,
D. Belitz, T.R. Kirkpatrick and R. Narayanan. This work was supported
in part by the DFG under grant No.
Vo659/2 and by the NSF under grant No. DMR--98--70597.
Part of this work was performed at the Aspen Center for Physics.

\vfill\eject
\end{document}